# Non-collinear Magnetic states of $Mn_5Ge_3$ compound

A. Stroppa[*] and M. Peressi

*Dipartimento di Fisica Teorica, Università di Trieste,*

*Strada Costiera 11, I-34014 Trieste, Italy and*

*CNR-INFM DEMOCRITOS National Simulation Center, Trieste, Italy*

## Abstract

$Mn_5Ge_3$ thin films epitaxially grown on Ge(111) exhibit metallic conductivity and strong ferromagnetism up to about 300 K. Recent experiments suggest a non-collinear spin structure. In order to gain deep insights into the magnetic structure of this compound, we have performed fully unconstrained ab–initio pseudopotential calculations within density functional theory, investigating the different magnetic states corresponding to Collinear (C) and Non-Collinear (NC) spin configurations. We focus on their relative stability under pressure and strain field. Under pressure, the C and NC configurations are degenerate, suggesting the possible occurrence of accidental magnetic degeneracy also in $Mn_5Ge_3$ real samples. We found a continuous transition from a ferromagnetic C low-spin state at small volumes to a NC high-spin state at higher volumes. Remarkably, the degeneracy is definitely removed under the effect of uniaxial strain: in particular, NC spin configuration is favoured under tensile uniaxial strain.

* Presently at: Institute of Material Physics, University of Vienna, Sensengasse 8/12, A-1090 Wien, Austria and Center for Computational Materials Science (CMS), Wien, Austria

# 1 Introduction

Among the Dilute Magnetic Semiconductors (DMS's) which are promising for spintronic applications, Mn-doped Ge has recently attracted much attention [1–3].

One of the problems in DMS is indeed the solubility of Mn in the semiconducting host: beyond a certain critical Mn concentration, a tendency towards clustering and phase separation occurs, thereby limiting the homogeneity and growth control that are required for materials to be used in spintronic applications. This tendency was observed also during Mn alloying of Ge samples: $Mn_xGe_y$ precipitates were detected during out-of-equilibrium growth [4].

Intermetallic compounds of Mn and Ge occur in several different stoichiometries and crystallographic phases [5]. Among them, $Mn_5Ge_3$ shows ferromagnetism with a Curie temperature of 300 K [6–8]. Recently, ferromagnetic $Mn_5Ge_3$ thin films grown epitaxially on Ge(111) by means of solid-phase epitaxy [8, 9] exhibited metallic conductivity and strong ferromagnetism up to 296 K, thus holding out promise for use in spin injection.

Very few first-principles calculations have been carried out for this compound [10] and all of them are limited within the framework of collinear spin-density functional theory (DFT), although recent experiments suggest the possibility for non-collinear magnetic state [11]. Therefore we have performed new DFT studies including the possibility of non-collinearity. We focused in particular on the relative stability of the Collinear (C) and Non-Collinear (NC) spin configurations under pressure and strain field and on the analysis of their magnetic properties. We have found that the C and NC self-consistent solutions, although corresponding to slightly different magnetic states, are degenerate over a wide range of external pressure within the numerical uncertainty. We also found a continuous transition from a ferromagnetic C low-spin state at small volumes to a ferromagnetic NC high-spin state at higher volumes. Interestingly, the degeneracy is definitely removed under the effect of uniaxial structural distortions.

The results are presented as follows: in Sect. 2 we report the computational approach along with the description of the crystal structure; in Sect. 3 we approach the study of the magnetic state, and we focus on the effect of the pressure in Sect. 4 and on the effect of the strain field in Sect. 5; finally, in Sect. 6 we draw our conclusions.

## 2  Structure and computational details

We performed a DFT study in the Local Spin Density Approximation (LSDA) by using state-of-the-art first-principles pseudopotential self-consistent calculations. The scheme of Ceperley and Adler [12] as parametrized by Perdew and Zunger [13] has been used for exchange-correlation functional. An ultrasoft pseudopotential [14] has been used for Mn atom, while norm-conserving PP has been considered for Ge. The kinetic energy cutoff has been fixed to 22 Ryd. A grid of (4,4,6) Monkhorst-Pack points has been used for integrations over the Brillouin zone. Test calculations have shown that with this choice of computational parameters the results are well converged, with an estimated error of the calculated atomic magnetic moments by ~0.02 $\mu_B$ and of their directions by a fraction of a degree. We neglected the spin-orbit interactions. The choice of the reference direction for the magnetization is of course arbitrary.

The intermetallic compound $Mn_5Ge_3$ has the hexagonal $D8_8$-type crystal structure with space group 193 or P63mcm. The experimental cell dimensions at room temperature are $a_{hex}$= 0.718 and c=0.505 nm. The hexagonal cell contains 10 Mn and 6 Ge atoms. The Mn atoms can be distinguished into two different sublattices, say $Mn_1$ and $Mn_2$, due to their different coordination. The atomic positions in the hexagonal cell, in internal crystal coordinates, are:

- 4 $Mn_1$ in : $\pm(\frac{1}{3},\frac{2}{3},0); \pm(\frac{2}{3},\frac{1}{3},\frac{1}{2})$

- 6 $Mn_2$ in: $\pm(x,0,\frac{1}{4}); \pm(0,x,\frac{1}{4}), \pm(-x,-x,\frac{1}{4})$ where x=0.239

- 6 Ge in: $\pm(x,0,\frac{1}{4}); \pm(0,x,\frac{1}{4}), \pm(-x,-x,\frac{1}{4})$ where x=0.603.

In Fig. 1, we show (a) top, (b) perspective, and (c) side view of the hexagonal $Mn_5Ge_3$ cell. Four atomic layers are stacked along the [0001] or z direction. As it can be seen from Fig. 1, there are two different atomic planes perpendicular to the z direction: the first one containing only $Mn_1$ atoms at z=0 and z=$\frac{1}{2}$c (equivalent by symmetry) which form an hexagonal two–dimensional lattice; the second contains $Mn_2$ and Ge atoms at z= $\frac{1}{4}$c and z=$\frac{3}{4}$c (equivalent by symmetry). The coordination of $Mn_1$ and $Mn_2$ is as following:

- $Mn_1$ has 2 $Mn_1$ and 6 $Mn_2$ nearest neighbors at 0.252 nm and 0.306 nm; 6 Ge nearest neighbors at 0.253 nm;
- $Mn_2$ has 2 $Mn_2$, 4 $Mn_2$, and 4 $Mn_1$ nearest neighbors at 0.298, 0.305 and 0.306 nm respectively; 2 Ge, 1 Ge, and 2 Ge nearest neighbors at 0.248, 0.261 and 0.276 nm.

## 3   Collinear versus non-collinear ferromagnetic state

The magnetization of the $Mn_5Ge_3$ compound is almost entirely due to the Mn atoms. Our calculations for the total magnetization give an average magnetic moment per Mn atom of ~2.5 $\mu_B$ which agree (within 0.1 $\mu_B$) with other DFT calculations [10, 11] and measurements on epitaxial [11] and bulk samples [6, 15].

The spatial distribution of the magnetization in $Mn_5Ge_3$ has been investigated already several years ago by J.B. Forsyth et al. [7] who reported magnetic moments of 1.96 and 3.23 $\mu_B$ for $Mn_1$ and $Mn_2$, with the $Mn_2$ atoms carrying the largest magnetic moment. The smaller magnetic moment of $Mn_1$ has been attributed to the direct Mn-Mn interaction which occurs at these very short distances. More recent measurements [11] suggest the picture of a metallic ferromagnet with a spin structure likely noncollinear. In order to gain insight in the magnetic structure of this compound, we have investigated both the ferromagnetic (FM) collinear (C) and NC state.

As observed in Ref. [16] for the similar compound $Mn_5Si_3$ (which has the same crystal structure as $Mn_5Ge_3$), even magnetic atoms with the same chemical environment can have different magnetic moments. The 4 $Mn_1$ atoms in the unit cell are coordinated with other $Mn_1$ atoms at a very short distance (see previous Section) and this fact can be responsible of peculiar magnetic properties of $Mn_1$ atoms [16]. Following this observation, we have released the constraint of spin collinearity and allowed for the possibility of different magnetic moments on the different atoms, even on those which are equivalent by symmetry.

In order to find the macroscopic equilibrium structure, we study for the C and NC state the variation of the Total Energy as a function of the unit cell volume, keeping the c/a ratio equal to the experimental value $((c/a)^{exp}=0.703)$. Results are shown in Fig. 2: the two curves overlap within the numerical error ( ~1 few meV) and the equilibrium volume is the same for the two magnetic structures, only slightly smaller than the experimental one ($a^{th}_{hex}$=0.698 nm, $a^{exp}_{hex}$=0.718 nm).

The two energetically degenerate states C and NC have also very similar global magnetic properties at the equilibrium volume. The calculated total(absolute) magnetizations are 25.01(26.90) and 24.90(26.90) $\mu_B$ per unit cell respectively for the C and NC cases, i.e. equal within the numerical uncertainty which is about 0.2 $\mu_B$/cell.

The difference between absolute and total magnetizations corresponds to the presence of region of negative contribution to the magnetization in the unit cell. We found an induced negative polarization on Ge atoms (see Tab. I) which are antiferromagnetically coupled with the Mn atoms. However, the induced magnetization on the Ge atoms is quite small with respect to the negative contribution to the total magnetization, thus suggesting a spatially diffuse contribution of negative polarity, along with the trends of the experimental observations [7].

The difference in the C and NC states is subtle and at the atomic level. In the NC case, the $Mn_1$ atoms do not exhibit the same magnetic moment, at variance with the $Mn_2$ atoms and at variance with the C case. They can be distinguished into two types, characterized by the following positions:

- $Mn_1\alpha$ in $(\frac{1}{3},\frac{2}{3},0)$; $-(\frac{2}{3},\frac{1}{3},\frac{1}{2})$

- $Mn_1\beta$ in $-(\frac{1}{3},\frac{2}{3},0)$; $(\frac{2}{3},\frac{1}{3},\frac{1}{2})$

and with magnetic moments equal in magnitude but with slightly different directions in the $\alpha$ and $\beta$ cases.

In Tab. I we report the details, showing the local (atomic) magnetic moments in polar and cartesian coordinates. In Fig. 3 we show the unit cell with the arrows representing the magnetic moments in magnitude and direction. Interestingly, neither $Mn_1\alpha$ nor $Mn_1\beta$ are fully aligned with $Mn_2$: the deviation from collinearity is small (with an angle of about $\pm 6°$) but definitely non zero. We want to remark that the calculated magnitude of the magnetic moments for $Mn_1\alpha$, $Mn_1\beta$ and $Mn_2$ are very close to the experimental data (with a small underestimate for $Mn_2$), which however do not provide any precise information about non-collinearity [7].

In conclusion, since we found that C and NC states have competing energies (accidental magnetic degeneracy) [17], we suggest that the occurrence of the NC state (or the coexistence of the NC and C phase) for $Mn_5Ge_3$ in real samples cannot be excluded a priori, although the deviation from the collinearity at the equilibrium volume is quite small. Two issues have still to be addressed, namely the effects of (i) isotropic pressure and (ii) anisotropic applied strain field. We discuss these points in the next Sections.

# 4 Effect of pressure

In this section we want to discuss the effect of pressure $p$ on the magnetic state, for $p>0$ ($p<0$), that is an isotropic compression (expansion) of the crystal unit cell, keeping the c/a ratio fixed to the experimental value. In Tab. II, we show the results for four different volumes, smaller (Low-Volume) and larger (High-Volume) than the equilibrium volume $V_0$: $V_1 < V_2 < V_0 < V_3 < V_4$ (corresponding to $a_{hex}$=0.635, 0.672, 0.741, 0.767 nm respectively). The Low-Volume state is characterized by a low value of the total and absolute magnetic moment in the unit cell ($M_{Tot}$ and $M_{Abs}$): we can characterize it as Low-Volume Low-Spin (LV-LS) magnetic state. On the other hand, the High-Volume state is characterized by high value of magnetization (total and absolute): we characterize this state as High-Volume High-Spin (HV-HS) magnetic state.

There is gradual change of the magnetic properties going from the LV-LS to HV-HS state. The atomic magnetic moments in the unit cell gradually increase but the $Mn_1$ atom type always carries the lowest magnetic moment. The $Mn_2$ atoms do not change the directions of their magnetic moments with volume. The same apply for Ge atoms which re-main nearly antiferromagnetically aligned with the Mn atoms. At variance, for $Mn_1^\alpha$ and $Mn_1^\beta$, the orientations of the magnetic moments change with volume. The different magnetic properties of $Mn_1$ and $Mn_2$ atoms would suggest that the two magnetic sublattices ($Mn_1$ and $Mn_2$) have different exchange coupling constants [11].

Interestingly, the LV-LS show a collinear coupling. As the volume increases, the $Mn_1$ spins catalyse the continuous transition from the ferromagnetic configuration for small lattice constant to a non-collinear alignment for larger lattice constants. We remark that this behaviour is opposite to what found in other systems, like Iron-Nickel alloys where ab–initio calculations have shown a continuous transition from the ferromagnetic state at high volumes to a non-collinear configuration at low volumes [18]. In Fig. 4 we show the calculated ground-state spin configurations for some selected atomic volumes. The figure shows the transition from LS-LV-C state to the HS-HV-NC state: as the volume increases the system evolves towards an increasingly disordered non-collinear state. The angle between $Mn_1^\alpha$ ($Mn_1^\beta$) and $Mn_2$ spins goes from 0° to a maximum of about 10°.

We conclude the section as follows: despite C and NC configurations are degenerate in energy over a wide range of pressures and are similar at equilibrium also in their magnetic properties, they differ quite a lot in the local magnetic properties as long as the volume increases.

There is a continuous transition from the a ferromagnetic collinear low-spin at small volumes to a ferromagnetic non-collinear high-spin at higher volumes. In next Section, we explore the possibility of removing the degeneracy with a non-isotropic strain field.

# 5 Effect of uniaxial distortions

In order to take into account the effect of uniaxial structural distortions on the magnetic properties, we have performed calculations varying the c/a ratio for some selected values of $a_{hex}$, namely $a_{hex}$=0.672, 0.698 (the equilibrium value at zero pressure), and 0.741 nm. Fig. 5 shows the total energy for the C and NC states as a function of c/a (each panel corresponds to a fixed value of $a_{hex}$). The curves clearly show that the accidental magnetic degeneracy is definitely removed by c/a ratios greater than the experimental value by about 20-30%. Furthermore, higher c/a ratio stabilizes the NC configuration. This is along the results of a recent paper [19] which reports that even in typical ferromagnetic materials (e.g., Fe, Co, and Ni) high pressure and/or strain conditions could stabilize a NC magnetic order.

# 6 Conclusions

We have shown that the magnetic structure of $Mn_5Ge_3$ intermetallic compound has two competing phases (accidental magnetic degeneracy), with collinear and non-collinear spin configurations. Although the deviation from collinearity at equilibrium is small, this could play a non-negligible role in real samples[11].

We have shown that the degeneracy is removed under uniaxial strain.

Further studies are in progress in order to clarify the details of the magnetic transition, its rationale in terms of underlying electronic and magnetic structure and possible experimentally detectable consequences.

# 7  Acknowledgments

The authors thank A. Debernardi for useful discussions. Figs 1, 3, 4 have been done using the XCrySDen program [20]. Computational resources have been obtained partly within the "Iniziativa Trasversale di Calcolo Parallelo" of the Italian Istituto Nazionale per la Fisica della Materia (INFM) and partly within the agreement between the University of Trieste and the Consorzio Interuniversitario CINECA (Italy).

TABLE I: Calculated atomic magnetic moments of $Mn_5Ge_3$ compound at equilibrium volume ($a_{hex}$= 0.698 nm, c/a=0.703) for the NC case, expressed in polar (decimal degrees for angles, $\mu_B$ for the intensity) and cartesian components ($\mu_B$).

| Atom type | $\mu(\mu_B)$ | $\vartheta(°)$ | $\varphi(°)$ | $\mu_x(\mu_B)$ | $\mu_y(\mu_B)$ | $\mu_z(\mu_B)$ |
|---|---|---|---|---|---|---|
| $M_{1\alpha}$ | 1.91 | 24 | 29 | 0.69 | 0.38 | 1.74 |
| $M_{1\beta}$ | 1.91 | 36 | 18 | 1.07 | 0.35 | 1.54 |
| $M_2$ | 2.87 | 30 | 22 | 1.33 | 0.54 | 2.48 |
| Ge | 0.09 | 150 | -158 | -0.04 | -0.03 | -0.08 |

TABLE II: Self-consistent magnetic spin configurations of the $Mn_5Ge_3$ compound at four different volumes. The atomic magnetic moments are expressed in polar components. The total ($M_{tot}$) and absolute ($M_{abs}$) magnetizations of the unit cell are also reported.

| $V_1 = 0.1559 nm^3$ | | | | $V_2 = 0.1849 nm^3$ | | | |
|---|---|---|---|---|---|---|---|
| $M_{tot} = 2.02\mu_B$  $M_{abs} = 2.13\mu_B$ | | | | $M_{tot} = 17.37\mu_B$  $M_{abs} = 18.70\mu_B$ | | | |
| Atom type | $|\mu|(\mu_B)$ | $\theta(°)$ | $\phi(°)$ | Atom type | $|\mu|(\mu_B)$ | $\theta(°)$ | $\phi(°)$ |
| $Mn1_\alpha$ | 0.22 | 30 | 22 | $Mn1_\alpha$ | 1.19 | 30 | 22 |
| $Mn1_\beta$ | 0.22 | 30 | 22 | $Mn1_\beta$ | 1.19 | 30 | 22 |
| $Mn_2$ | 0.17 | 30 | 22 | $Mn_2$ | 2.08 | 30 | 22 |
| Ge | 0.01 | 149 | -157 | Ge | 0.06 | 149 | -157 |

| $V_3 = 0.2476 nm^3$ | | | | $V_4 = 0.2749 nm^3$ | | | |
|---|---|---|---|---|---|---|---|
| $M_{tot} = 28.72\mu_B$  $M_{abs} = 31.65\mu_B$ | | | | $M_{tot} = 30.90\mu_B$  $M_{abs} = 34.21\mu_B$ | | | |
| Atom type | $|\mu|(\mu_B)$ | $\theta(°)$ | $\phi(°)$ | Atom type | $|\mu|(\mu_B)$ | $\theta(°)$ | $\phi(°)$ |
| $Mn1_\alpha$ | 2.57 | 21 | 34 | $Mn1_\alpha$ | 2.96 | 20 | 37 |
| $Mn1_\beta$ | 2.57 | 41 | 15 | $Mn1_\beta$ | 2.96 | 43 | 14 |
| $Mn_2$ | 3.13 | 30 | 22 | $Mn_2$ | 3.23 | 30 | 22 |
| Ge | 0.14 | 149 | -157 | Ge | 0.14 | 149 | -157 |

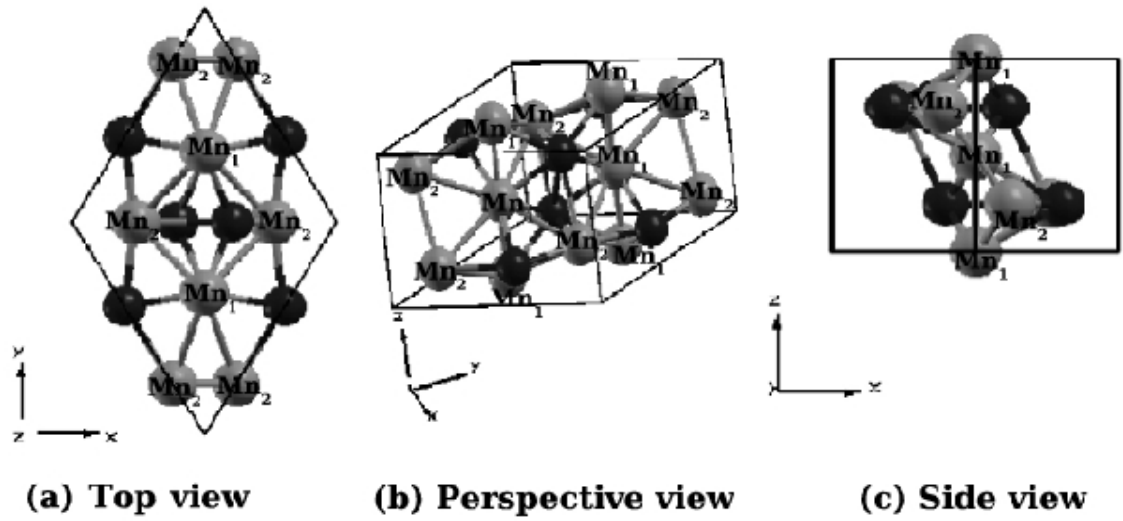

FIG. 1: Unit cell of $Mn_5Ge_3$: top (a), perspective (b) and (c) side view. $Mn_1$ and $Mn_2$ label the two types of Mn atoms in the unit cell.

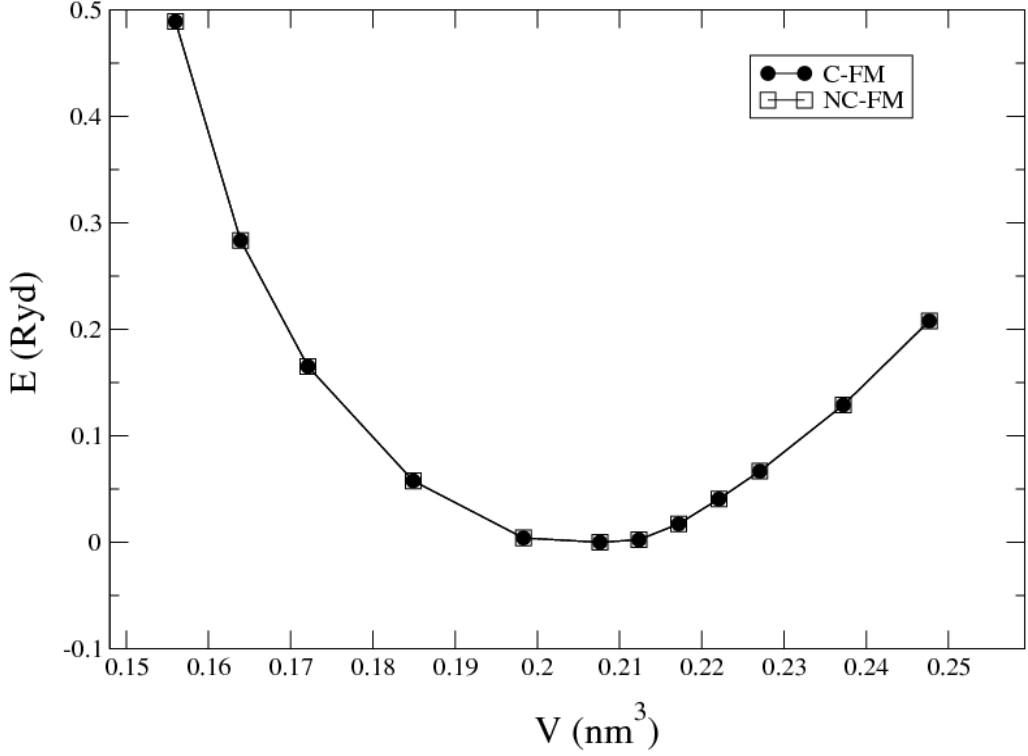

FIG. 2: Total energy as a function of the volume of the hexagonal unit cell keeping the c/a ratio equal to the experimental value ((c/a)$^{exp}$=0.703) for the collinear (C) and non-collinear (NC) ferromagnetic (FM) state. Energies have been rescaled with respect to the ground state energy of the C-FM state. The two curves are degenerate within the numerical error, estimated to be about few meV.

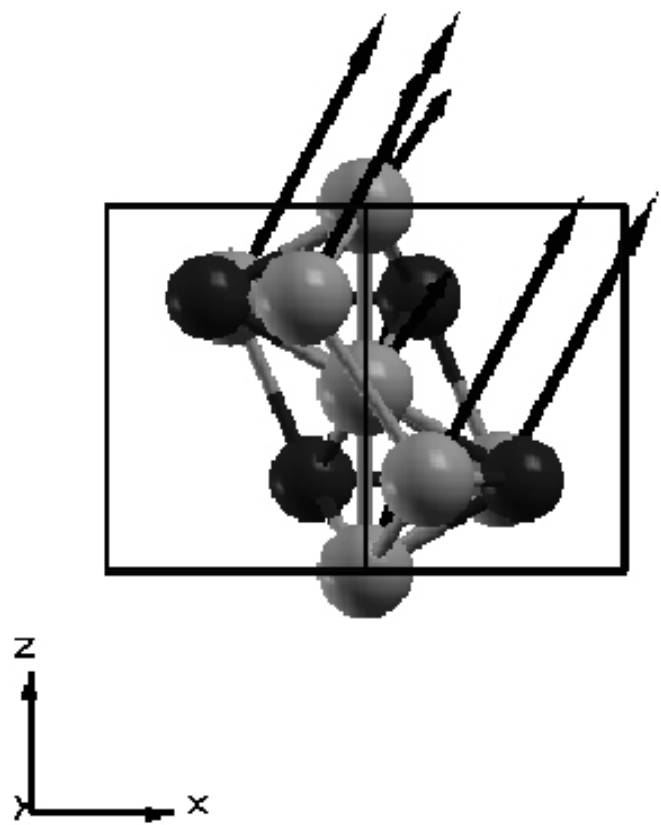

FIG. 3: Side view of $Mn_5Ge_3$ compound with the magnetic moments on Mn atoms. The length of the arrows is proportional to the intensity. The Ge moments are not reported. Black (grey) spheres represent Ge(Mn) atoms. It is evident the small deviations from collinearity of the magnetic moments.

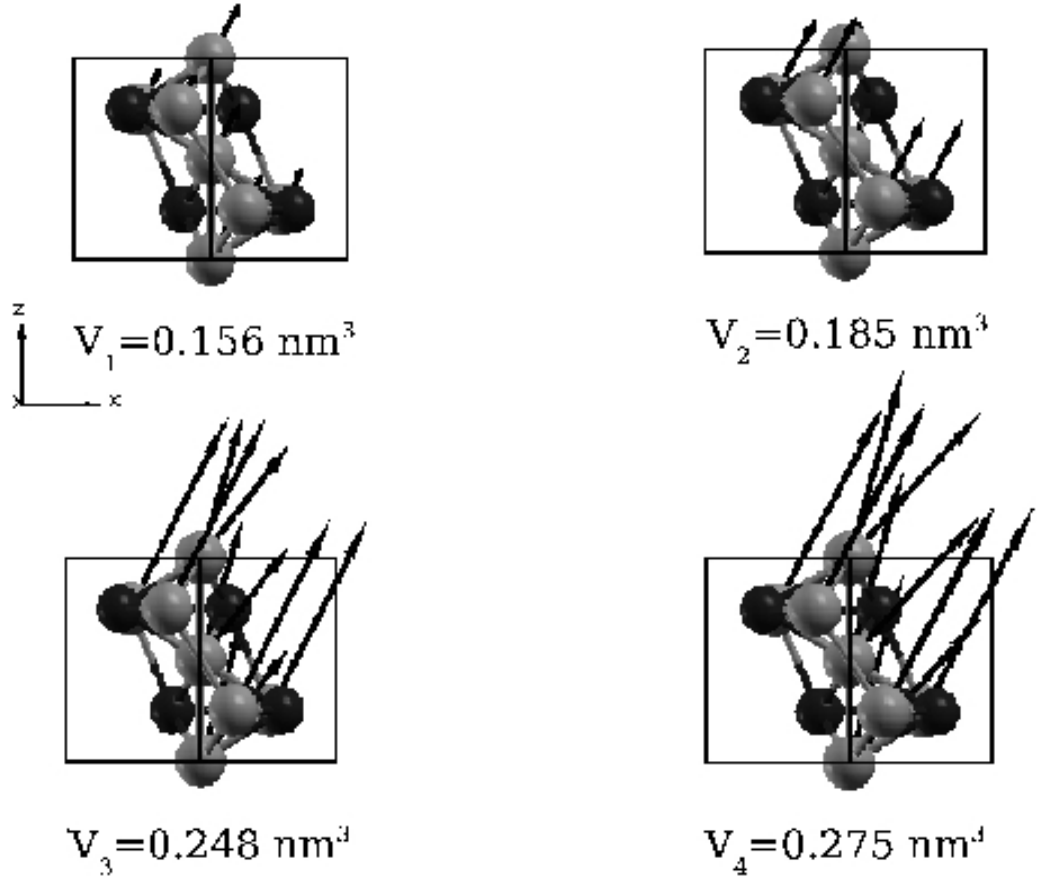

FIG. 4: Side view of $Mn_5Ge_3$ compound for different volume corresponding to different values of $a_{hex}$ (keeping the c/a ratio equal to the experimental value). The arrows correspond to the magnetic moments on Mn atoms (in scale for different $a_{hex}$, except for $a_{hex}$=0.635 nm, where they have been multiplied by a factor of 60). The Ge moments are not shown. Black (grey) spheres represent Ge(Mn) atoms. It is evident the deviation from collinearity (C) going from the Low-Volume Low-Spin (LS-LV) to High-Volume High Spin (HS-HV) state (see text).

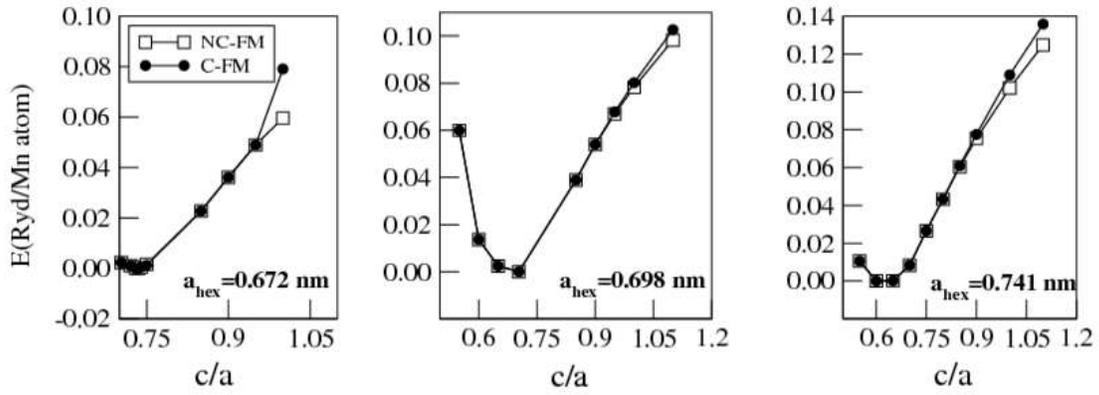

FIG. 5: Total energy for the Collinear (circles) and Non-Collinear (squares) spin configurations as a function of the c/a ratio for some selected values of $a_{hex}$, namely $a_{hex}$=0.672, 0.698 (the equilibrium value at zero pressure), and 0.741 nm. The energy minima are arbitrarily set to zero.